\documentstyle{article}
\input math_macros

\font\tenrm=cmr10
\font\tenit=cmti10
\font\elevenbf=cmbx10 scaled\magstep 1
\font\elevenrm=cmr10 scaled\magstep 1
\font\elevenit=cmti10 scaled\magstep 1

\renewenvironment{thebibliography}[1]
 { \elevenrm
   \begin{list}{\arabic{enumi}.}
    {\usecounter{enumi} \setlength{\parsep}{0pt}
     \setlength{\itemsep}{3pt} \settowidth{\labelwidth}{#1.}
     \sloppy
    }}{\end{list}}

\begin{document}
\begin{center}{
               GLOBAL CONSTRAINTS OF GAUSS' LAW AND THE\\
               \vglue 3pt
                 SOLUTION TO THE STRONG CP PROBLEM\\}
\vglue 5pt
\vglue 1.0cm
{\tenrm HUAZHONG ZHANG \\}
\baselineskip=13pt
{\tenit P.O.Box 17660, Jackson State
University\\}
\baselineskip=12pt
{\tenit Jackson, MS 39217, USA\\}
\vglue 0.3cm
\vglue 0.8cm
{\tenrm ABSTRACT}
\end{center}
\vglue 0.3cm
{\rightskip=3pc
 \leftskip=3pc
 \tenrm\baselineskip=12pt
 \noindent
This is a brief review on the work done recently.
It is shown that the global constraints of Gauss' law ensure that
the vacuum angle must be quantized in gauge theories with magnetic
monopoles. Our quantization rule is given as
$\theta=0$, or $\theta=2\pi N/n~ (n\neq 0)$ with integer n being the
relevant topological charge of the magnetic monopole and N is an unfixed
integer in this approach. This theoretically confirms further
the conclusion originally proposed by the author$^1$ that the strong CP
problem can be solved due to the existence of magnetic monopoles, the fact
that the strong CP-violation can be only so small or vanishing may imply
the existence of magnetic monopoles, and the universe is open.
\vglue 0.6cm}
{\elevenbf\noindent 1. Introduction and the Summary of the Main Results}
\vglue 0.4cm
\elevenrm
It is well known that the effective $\theta$ term in QCD with an arbitrary
effective and chiral invariant
vacuum angle $\theta+arg(detM_{quark})$, which will be denoted as
$\theta$ hereafter can induce CP violations in strong interactions.
However, the possible electric dipole moment of the netron puts a limit
$\theta\leq 10^{-9}$ (modulo 2$\pi$). The need for understanding this
leads to the strong CP problem.
Recently, a non-perturbative solution to the Strong CP problem with
magnetic monopoles has been proposed originally by the author$^{1,2}$.
This may be interesting since the axions needed in the well-known
Peccei-Quinn solution with an additional U(1) symmetry$^3$ have not
been observed. In our solution$^{1,2}$, it is proposed that the vacuum
angle with magnetic monopoles must be quantized. Our
quantization rule is derived by two different methods. This is given
by $\theta=0$, or $\theta=2\pi N/n~ (n\neq 0)$ with integer n being the
relevant topological charge of the magnetic monopole and N is an integer.
In Ref. 1, it is shown that the integer N can be chosen as 1. The first
method$^1$ is to show the existence of a monopole structure in the
relevant gauge orbit space in Scherodinger formulation$^{4,5}$, and using
the Dirac quantization rule for having a well-defined wave funtional.
The relevance to the $U_A(1)$ problem and the relevant vortex structures
are also discussed in Ref.1.
The second method is to show that there exist well-defined gauge
transformations which will ensure the quantization of $\theta$ by the
constraints of Gauss's law due to the non-abelican electric charges
carried by the magnetic monopoles proportional to $\theta$ as noted
in Ref. 6 and generalized in Ref. 7.

Therefore, we conclude that strong CP problem can be solved due to
the quantization of $\theta$ in the presence of magnetic monopoles,
for example monopoles of topological charge $n=\pm 1$ with
$\theta=\pm 2\pi$, or $n\geq 2\pi 10^9$ with $\theta\leq 10^{-9}$.
Moreover, the existence of non-vanishing magnetic flux
through the space boundary implies that the universe must be open.
In this note, we will briefly discuss and review the method in Ref. 2.

\vglue 0.6cm
{\elevenbf\noindent 2. Constraints of Gauss' Law and the Qantization
Rule for $\theta$}
\vglue 0.4cm
In the Schrodinger formulation, the gauge theory system is described by
a Hamiltonian equation with the constraints of Gauss' law for the wave
functional $\Psi [A]$.
Consider the non-abelian gauge theories with a $\theta$ term
with a non-singular magnetic monopole at the origin.
Such a $\theta$ term in the presence of magnetic monopole was first
considered$^{8}$ relevant to $U_A(1)$ problem. It was noted by Witten and
generalized in Ref. 7 that monopoles of magnetic charges $q=G_aL_a$,
can carry induced non-abelian electric charges
$q_e=m\frac{q}{2\pi}-\frac{\theta q}{4\pi^2},~ m\in Z,$
where $L_a$ (a=1, 2,... dim(G)) form a basis for the Lie algebra of the
gauge group. For an arbitrary value of $\theta$, monopoles must be dyons.
For simplicity, we will consider the case of static field
for our purpose.
The $\theta$ term can be written as
\begin{equation}
{\cal L}_{\theta}=\frac{\theta}{4\pi^2}{\bf E}\cdot{\bf B}.
\end{equation}
Where
${\bf E}^a=-{\bf {\nabla}}\phi^a-C_{abc}{\bf A}^b\phi^c,~~
{\bf B}^a={\bf {\nabla}}\times{\bf A}^a
-\frac{1}{2}C_{abc}{\bf A}^b\times{\bf A}^c$
with $\phi^a={A_0}^a$, and $C_{abc}$ being the structure constants
of the Lie algebra.
With integration by part, the $\theta$ term
induces a non-abelian electric charge density
\begin{equation}
\rho^a=\frac{\theta}{4\pi^2}{\bf {\nabla}}\cdot {\bf B}^a
-\frac{\theta}{4\pi^2}C_{abc}{\bf A}^b\cdot {\bf B}^c.
\end{equation}
The first and the second terms will induce a non-abelian electric
charge at the monopole$^{6,7}$ and a non-abelian electric charge density in
the space outside the monopole respectively$^{7}$.

We will choose the Weyl gauge $A_0=0$. The constraints of
Gauss' law for ${\Psi}_{phys}[{\bf A}]$ is given by
\begin{equation}
\{-i{D_i}^{ab}\frac{\delta}{\delta
{A_i}^b}-\rho^a\}{\Psi}_{phys}[{\bf A}]=0,
\end{equation}
where ${D_i}^{ab}=\delta^{ab}\partial_i+C_{abc}{A_i}^c$.
Now multiply the above equation for the constraints of Gauss law by
infinitesimal
$\lambda^a({\bf x})$, sum over a and integrate over ${\bf x}$. Then
we obtain
\begin{equation}
\int d^3x\{i(D_i\lambda)^a\frac{\delta}{\delta
{A_i}^a}-\lambda^a\rho^a\}
{\Psi}_{phys}[{\bf A}]=0.
\end{equation}
This gives
$\delta_{\lambda}{\Psi}_{phys}[{\bf A}]=\{-i\int
d^3x\lambda^a\rho^a\}
{\Psi}_{phys}[{\bf A}]$. For a well-defined finite gauge
transformation with gauge function
$g({\bf x})=exp\{\lambda^a({\bf x})L_a\}$, we can consider the path
$g({\bf x},t)=exp\{t\lambda^a({\bf x}) L_a\}~~~t\in [0,1]$,
connecting $g({\bf x})$ and the identity element e.
Consider a closed loop $C^1$ with
$g({\bf x},1)=g({\bf x},0)=e$, the state wave functional
${\Psi}_{phys}[{\bf A}]$ is invariant.
Then by integration along the closed path$^4$,
with ${\bf A}^g=g^{-1}{\bf A}g+g^{-1}{\bf\nabla}g$ and
$tr(L_aL_b)=-\frac{1}{2}\delta_{ab}$, in terms of field strength
two form F, this gives$^2$ topologically
\begin{equation}
\frac{\theta}{2\pi^2}
tr\{\int_{S^2}F\int_{C^1}\delta gg^{-1}\}=2\pi k,
\end{equation}
with k being integers.
Now the magnetic charges can be transformed into a given Cartan subalgebra
with the basis $H_i$ (i=1,2...r) as
\begin{equation}
\int_{S^2}F=G_0=4\pi\sum_{i=1}^{r}\beta^{i}H_{i}.
\end{equation}
The constraints from the well-defined gauge
transformations must be physical. Consider the following$^9$ well-defined
U(1) gauge subgroup on the space boundary given by
\begin{equation}
g(t)=exp\{4\pi
mt\sum_{i,j}\frac{(\alpha_i)^jH_j}{<\alpha_i,\alpha_i>}\},~m\in Z,
{}~t\in [0,1],
\end{equation}
where the $\alpha_i$ (i=1,2,...,r) are the simple roots of the Lie algebra.
We obtain $exp\{i\frac{\theta nm}{2\pi}\}=1$.
Where n is the effective topological number of the magnetic
monopole defined as$^1$
\begin{equation}
n=-2<\delta',\beta>
=-2\sum_{i=1}^{r}\frac{<\alpha_i,\beta>}{<\alpha_i,\alpha_i>}.
\end{equation}
Therefore, we obtain the quantization rule
\begin{equation}
\theta=\frac{2\pi N}{n},~~(n\neq 0),
\end{equation}
with N being an integer.

\vglue 0.5cm
{\elevenbf\noindent 3. References \hfil}
\vglue 0.4cm

\end{document}